\documentclass{iau} 
\usepackage{graphicx}
\usepackage{aas_macros,natbib}

\title[The Impact of Binary Models] 
{Interpreting Galaxy Properties with Improved Modelling}

\author[E.~R.~Stanway \& J.~J.~Eldridge]   
{E.~R.~Stanway$^1$
 \and J.~J.~Eldridge$^2$}

\affiliation{$^1$Dept. of Physics, University of Warwick, Gibbet Hill Road, Coventry, CV7 4AL, UK \\ email: {\tt e.r.stanway@warwick.ac.uk} \\[\affilskip]
$^2$Dept. of Physics, University of Auckland, Private Bag 92019, Auckland, New Zealand \\email: {\tt j.eldridge@auckland.ac.nz}}

\pubyear{2019}
\volume{352}  
\jname{Uncovering Early Galaxy Evolution in the ALMA and JWST Era}
\editors{E. da Cunha \& J. Hodges, eds.}
\begin{document}

\maketitle

\begin{abstract}
  Observations of star-forming galaxies in the distant Universe  have confirmed the importance of massive stars in shaping galaxy emission and evolution. Distant stellar populations are unresolved, and the limited data available must be interpreted in the context of stellar population models. Understanding these populations, and their evolution with age and heavy element content  is  key to interpreting processes such as supernovae, cosmic reionization and the chemical enrichment of the Universe. With the upcoming launch of JWST and observations of galaxies within a billion years of the Big Bang, the uncertainties in modelling massive stars -  particularly their interactions with binary companions - are becoming increasingly important to our interpretation of the high redshift Universe. In turn, observations of distant stellar populations provide ever stronger tests against which to gauge the success of, and flaws in, current massive star models. Here we briefly review the current status binary stellar population synthesis.
  
\end{abstract}

\firstsection 
\section{Background}

The galaxies observed in the distant Universe ($z>2$) differ in
significant respects from those seen more locally. It has been known
for many years that the typical star formation rates of massive
galaxies were much higher in the past than at the current time \citep[e.g.][]{1996MNRAS.283.1388M}, and
the typical stellar populations rather younger. Since the Universe is
enriched by supernovae and winds from old stars, it
also follows that the average metallicity of galaxies decreases as one
moves towards higher redshifts.

The star-forming nature of typical galaxies in the distant Universe
has practical advantages. While the optical emission from these
sources is redshifted beyond the capabilities of most ground-based
instrumentation, star-forming galaxies are ultraviolet luminous. This
has allowed rest-frame ultraviolet (rest-UV) observations to identify
large numbers of photometrically selected galaxy candidates at high
redshift, and also enabled spectroscopic confirmation or even
characterisation of a significant fraction of these based on
ultraviolet emission and absorption features.

The rest-UV emission of star-forming galaxies at high redshift, and
indeed at any redshift, is entirely dominated by the most massive,
hottest and hence most UV-luminous and shortest-lived stars in a
stellar population. These emit the stellar continuum shortwards of
the Balmer break, and are also responsible for powering nebular
emission from photoionized H\,II regions. These are impossible to
resolve in the very distant Universe, but their presence can be
inferred from the strength of narrow (low velocity) emission features
in both rest-UV and rest-optical integrated light spectra of
unresolved sources.

In addition to the indirect evidence from the continuum and nebular
emission, certain key features have been identified as diagnostic of
the presence of massive stars at high redshift. Prominent amongst
these is the He\,II 1640\AA\ emission line which occurs both in
stacked spectra of typical ultraviolet-selected galaxies, and in
individual sources where sufficiently bright. While a subset of these
appear to show narrow lines \citep[suggesting an emission source
  embedded in highly ionized nebular gas,
  e.g.][]{2010ApJ...719.1168E}, many detected He\,II lines appear to
be velocity broadened \citep[e.g.][]{2003ApJ...588...65S} - a feature
associated with the strong winds of massive, stripped-atmosphere
Wolf-Rayet stars in the local Universe.

Such massive star dominated, young stellar populations are rare in the
local Universe. High mass Galactic star forming regions typically lie
in the Galactic plane and are often heavily obscured by dust, making
observations (particularly in the rest-UV) challenging. They tend to
be limited in size, forming stars at a very different rate to those
seen in the distant population. Perhaps more importantly, very little
local star formation occcurs in regions less metal rich than the
Sun. The typical metallicities  expected in distant galaxies
(Z$<$0.2\,Z$_\odot$) are not represented in the resolved, Galactic
star forming population.

Amongst systems sufficiently close to resolve and analyse the
contribution of individual stars to the integrated light, perhaps the
closest analog to a distant galaxy can be seen in the 30\,Doradus
region of the large Magellanic Cloud. This super-starbust has been
subject to intense study in recent years and has revealed abundant
populations of massive stars \citep[e.g.][]{2018A&A...618A..73S,2018Sci...359...69S}, including `Very Massive Stars' (with
initial masses extending up to almost 300\,M$_\odot$), and a
near-ubiquitous population of stellar binaries \citep[e.g.][]{2012Sci...337..444S}. Indeed, the average
number of companions for massive stars
(M$_\mathrm{init}$$>$30\,M$_\odot$) actually exceeds 1, suggesting
significant numbers of triple and higher order multiple systems. It
has been estimated that 70\% of such massive stars will interact with
a companion during their lifetime
\citep[][]{2012Sci...337..444S}. Such interactions can include
mass transfer through Roche lobe overflow, stripping of atmospheres,
spin-orbit angular momentum transfer and, in extreme cases,
merger. All of these have the potential to modify the spectrum of
affected stars, and so will impact the interpretation of light from an
integrated stellar population.

Such considerations have resulted in a growing recognition
that the stellar population synthesis models widely used to interpret
galaxy properties in the local Universe require modification to
confront the very different physical conditions more commonly
encountered at high redshift. It has also highlighted the considerable
uncertainties that remain in massive star evolution, particularly at
low metallicities where local exemplars are rare or non-existant, and
when binary interaction effects are included. As ALMA opens new
windows on the infrared properties of distant galaxies, and JWST
promises unprecedented sensitivity to both the rest-UV and
rest-optical emission of distant sources, the need to explore these
uncertainties and improve our modelling of stellar populations is
becoming acute.

\section{Population and Spectral Synthesis}

When comparing observations to galaxy models, care must be taken to
account for the contributions of stellar types which may not have an
obvious impact on the observed integrated light, but which nonetheless
contribute to the physical interpretation. A widely-used approach to
doing this is the technique of stellar population synthesis.

\begin{figure}[tb]
\begin{center}
 \includegraphics[width=5.5in]{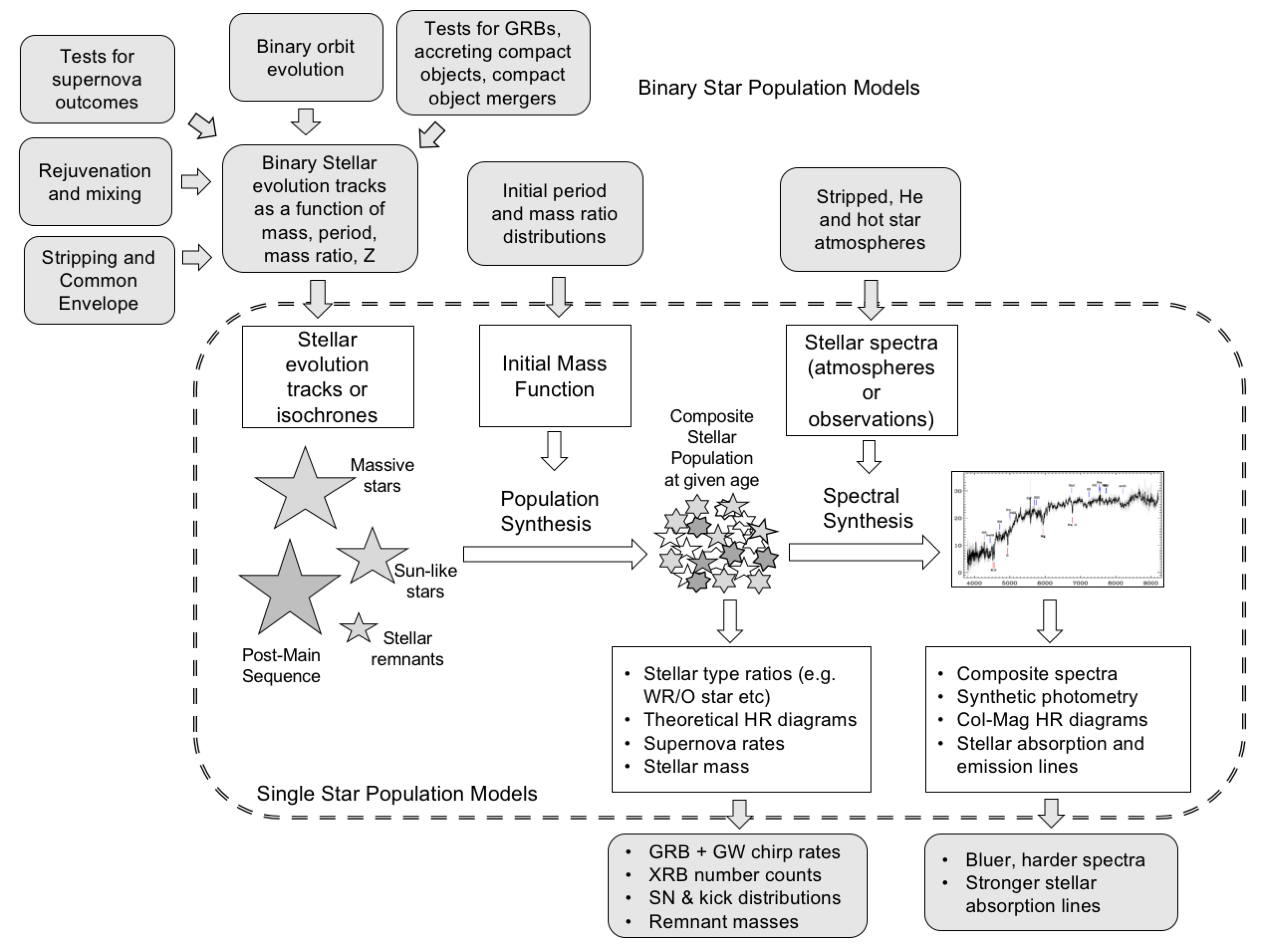} 
 \caption{A schematic diagram indicating the key elements, inputs and outputs of stellar population synthesis models, and how these are modified when binary stellar evolution pathways are considered.}
   \label{fig:popsynth}
\end{center}
\end{figure}

The key elements of this technique is illustrated schematically in the
central region of figure \ref{fig:popsynth} (within the dashed
line). The evolution of individual stars is assumed to be a well
understood function of initial stellar mass alone. Stellar evolution
models (occasionally combined with empirical approximations) are used
to establish the temperature, luminosity and perhaps surface gravity
of a star as a function of its age. Alternately isochrones, which give
the position of stars in luminosity-temperature space at fixed age as
a function of initial mass, can be substituted. These stellar models
are then combined, and the relative contibution of stars with a given
initial mass determined by an initial mass function (IMF). This
accounts for the presence of many low mass stars, as well as the
relatively few luminous and massive stars.

Such a population synthesis models can yield information on the
expected ratio of different stellar types as a function of age, as
well as on the number of massive stars reaching the end of their
lifetimes (i.e. type II core-collapse supernova rates), and
luminosity-temperature Hertzsprung-Russell diagrams.
Often comparison of models with data requires a further step. Spectral
synthesis models assign a theoretical atmosphere or empirically
derived spectrum to each modelled star on the basis of its physical
properties. These are summed with the appropriate weightings
determined by the population synthesis, to produce the integrated
light spectum of an entire stellar population as a function of age, as
well as synthetic photometry in desired wavebands.

The basic output of such models is known as a simple stellar
population or SSP
\citep[e.g.][]{2003MNRAS.344.1000B,2005MNRAS.362..799M,1999ApJS..123....3L,2014ApJS..212...14L}. It
estimates the properties of a group of stars sufficiently large to
fully sample the IMF, which formed simultaneously and instantaneously
at a known time before the point of observation. These SSPs can be
combined with a star formation history to create a complex or
composite stellar population (CSP) and further processed using
radiative transfer calculations to account for modifications to the
stellar spectrum by dust or gas, either in the circumstellar medium or
along the line of sight
\citep[e.g.][]{1998PASP..110..761F,2008MNRAS.388.1595D}. To reproduce
the characteristics of unresolved stellar populations, star formation
history, stellar properties, gas properties and dust extinction or
emission characteristics may need to be fit to the data
simultaneously.

\section{Binary Population Models}

Binary population synthesis models have been developed by the stellar physics community over a considerable timespan \citep[e.g.][]{1998A&A...334...21V,2000A&A...358..462V,2002MNRAS.329..897H,2004NewAR..48..861D,2004A&A...419.1057W,2008ApJS..174..223B,2013A&A...557A..87T}. The majority of these have focussed on the effects of binary interactions on massive stars, since these are both more likely to be found in a binary, and also show the most dramatic effects of binary interactions. However their use has been restricted to a narrow range of stellar evolution studies until recently, due both to the added computational burden they introduce and the additional assumptions and uncertainties they highlight.

Incorporating binary evolution pathways in a synthesis model is challenging. In the regions of Figure \ref{fig:popsynth} outside the dashed lines, we illustrated key elements that must be considered above and beyond those in a standard single-star population synthesis. Binary stellar evolution tracks are no longer a simple function of mass, but also depend on the mass ratio between primary and secondary and the initial binary separation. Evolution in the binary orbit and Roche lobes must be tracked as the stars age and interact. In addition to solving the equations of stellar structure and hydrostatic equilibrium for each layer of the star at each time step, mass transfer either to a companion or out of the system entirely must be calculated. This can lead to substantial changes in the surface composition of the stars (e.g. stripping), to rejuvenation of the star through rotational spin-up and subsequent mixing, and in extreme cases to phases of common envelope evolution. At the end of each stellar core-burning lifetime, the system must be tested for the occurrence of supernovae which may dissociate the binary.

Given the wide range of initial masses, mass-dependent binary fractions, periods and mass ratios represented in a stellar population, the number of individual models contributing to an output spectrum increases from perhaps hundreds to high thousands, and each model incorporates more steps with more physical parameters and effects to be calculated and considered. A range of approaches exist to address these difficulties.

The strategy adopted by some models, known as rapid-pop-synth models, is to take a semi-analytic approach. These sample the parameter spaces occupied by binaries with detailed models, and fit analytic approximations to the behaviour of key parameters at intermediate masses and separations \citep{2002MNRAS.329..897H,1995MNRAS.274..964P}. A strong advantage of such approaches is that a population can be rapidly constructed by integrating the analytic formula with appropriate weighting factors rather than by combining individual models. A disadvantage is that such approaches risk neglecting  the impact of stellar populations which, while having a significant impact on the integrated light at a given age or metallicity, may be small in number or occupy a narrow region of parameter space. Given that many of the responses to binary interactions are very sensitive to initial conditions, this is a concern when interpreting extreme or rare objects.

An alternate strategy is to work with a fixed grid of detailed (usually 1D) stellar evolution models which is sufficiently finely sampled that interpolation between grid elements is not required \citep[e.g.][]{2007MNRAS.380.1098H,2005MNRAS.364..503Z,2017PASA...34...58E}. Since these models contain information on the structure, mass transfer rates, composition and other physical properties of individual stars, more potentially observable properties of a population can be calculated (e.g. compact remnant mass distributions and accreting binary number counts, mass ranges for stripped stars etc) but at a computational cost. Such detailed models are relatively slow and resource-intensive to calculate.

When a binary population has been synthesised (incorporating assumptions for the binary star occurance fraction, the initial mass ratio and separation distributions) the individual stellar models can be wedded to atmosphere models in a spectral synthesis step as before. This often requires a wider range of atmospheres (incorporating, for example, stripped star and extremely hot massive stars) than in the single star population synthesis case. Binary models yield a wider range of output products which can be compared to observational constraints. Like any model, they will also only be as good as the assumptions made when they are calculated.

\section{Uncertainties in Population Synthesis}

A wide range of assumptions must be made in calculating spectral synthesis models of stellar populations. Most of these are discussed extensively in the literature \citep[see e.g.][]{2013ARA&A..51..393C,2017PASA...34...58E} but we summarise key points below.

\subsection{Uncertainties affecting all population synthesis}

\subsubsection*{Initial mass function}
The presumption of an initial mass function presents both practical and philisophical problems as discussed in detail by \citet{2018PASA...35...39H}. There is increasing evidence that a single, universal stellar mass function that describes the fragmentation pattern of all molecular gas clouds cannot explain the observed stellar populations in different galaxies. Reasons for variation may include the composition of the molecular clouds and their temperature profile. Since the initial mass function determines the ratio of the stars dominating the emission (i.e. UV luminous, massive stars in the high redshift case) to those which are difficult to observe (e.g. low mass stars which may contribute significantly to the rest-infrared), this has consequences for determining parameters such as the total stellar mass of a population, its absorption line indices or its population of compact remnants at late times.

\subsubsection*{Post-main sequence evolution}
Often overlooked inputs to the stellar models on which all population synthesis codes are built is the prescription adopted for the opacity of the stellar atmospheres and for stellar winds. These becomes particularly significant at high metallicities and during post-main sequence evolution. ALMA observations have provided direct evidence for the superwinds driven by asymptotic giant branch stars \citep[e.g.][]{2019arXiv190803311N}, but the mass loss rates of individual stars show a wide variation and it remains unclear which prescription for this best describes the average behaviour of stars, and how that depends on mass and other physical properties \citep[e.g.][]{2014ApJ...790...22R}. While the AGB stars which dominate old stellar populations are unlikely to be important in the distant Universe, massive stars evolve very rapidly and their behaviour on the giant branch, while short-lived, may include superwind phases which substantially alter their evolution.

\subsubsection*{Short-lived instability effects}
The calculation of stellar evolution models requires an inevitable compromise between resolution in both stellar evolution timescale and mass or radius, and the computational time required. A model which follows the evolution of a star though a Hubble time cannot probe the stellar properties of that star on timescales of hours, weeks, months or even a few years without being prohibitively expensive. While most codes are able to adjust their time steps adaptively, spanning a very large dynamic range in time tends to cause numerical instability. As a result, most evolution tracks fail to correctly follow very rapid, large magnitude changes in luminosity or mass. Examples of this include thermal pulsations of stars occurring in the instability strip, or rapid changes in internal structure such as occur at the onset of core burning for Helium, Carbon etc. Analytic or algorithmic prescriptions are used to bridge these phases of rapid evolution, but their impact on the integrated light of a population may be under or overestimated \citep[see e.g.][]{2005MNRAS.362..799M}.

\subsubsection*{Extrapolation to low metallicity}
As discussed in the Introduction, most of our empirical evidence for the evolution of individual stars is limited to the very local Universe, where massive stars forming at low metallicities are rare. The Extremely Metal Poor stars found in the Milky Way are very old, and therefore low mass, remnants of early star formation, and can provide little insight into processes which only occur in high mass stars. While theoretical stellar evolution models are largely based on well-understood physical principles valid at all metallicities, they are routinely cross checked and key assumptions calibrated based on such local observations. Extrapolations to significantly sub-Solar (and particularly any extremely-low Population III-like) metallicity are therefore often unsupported by observations.

\subsubsection*{Rotation}
Very few stellar evolution tracks exist for rotating stars. There is fair evidence that most massive stars are born rotating at an appreciable fraction of their break-up speed \citep[see e.g.][and references therein]{2019A&A...626A..50D}. At high metallicities, stars rapidly spin down due to stellar wind-driven mass and angular momentum loss. However it is possible that some systems continue to rotate for an appreciable fraction of their evolutionary lifetime. This induces additional mixing of the stellar interior, feeding additional light elements to the core, and mixing heavy elements into the stellar atmosphere. As a result the star burns both brighter and hotter. In the extreme case, a star can evolve as a chemically homogenous system (i.e. with uniform composition) rather than through the standard shell burning prescription. The result is a hotter, bluer spectrum than is obtained for non-rotating stars \citep[See e.g.][]{2015ApJ...800...97T,2014ApJS..212...14L}.

\subsubsection*{Extrapolation to the far ultraviolet}
The ionizing emission spectrum of any source is extremely difficult, if not impossible, to determine observationally. Neutral hydrogen in the interstellar and integalactic medium efficiently scatters hard ultraviolet and X-ray photons, while any trace of dust reprocesses the emission into the thermal infrared. The hardest radiation (extreme X-ray/gamma-ray) has such a long mean-free path that while it escapes absorption by hydrogen, it also has little impact on its surroundings. In the local Universe, it is possible to probe directly to a little shortwards of the hydrogen ionization edge at 912\AA, but the far-ultraviolet remains out of reach even here. As a result, the ionizing emission spectrum of stars of a given temperature, composition and structure is usually calculated from theory \citep[e.g.][]{2012A&A...538A..40G,2018A&A...615A..78G} and may be calibrated against the spectrum of ultraviolet and optical emission lines that results from reprocessing by nebular gas. Such calibration is dependent on the physical conditions of the gas, and again extrapolation to low metallicity is challenging.

\subsubsection*{Uncertainties in post-processing}
Each stage of post-processing required before comparison of simple stellar populations to observations (e.g. star formation history, nebular gas, dust etc) requires assumptions regarding the physical conditions in a given galaxy and introduces unavoidable degeneracies in the interpretation of any integrated light source. In some cases, these may be broken by the combination of detailed data across a wide wavelength range.

\subsection{Uncertainties specific to binary models}

\subsubsection*{Binary parameters}
Key inputs to a binary population synthesis are the fraction of stars in binaries, their initial period distribution as a function of initial mass, and the distribution of initial mass ratios as a function of both mass and period. These distributions have only recently been empirically determined for stellar populations in the local Universe \citep{2017ApJS..230...15M}. Inevitably these are derived from stellar populations which are at near-Solar metallicity, and also not at zero age. This is an area in its observational infancy, where much higher sensitivities and angular resolutions will be required to make substantial progress. There is some evidence that the binary fraction may be a sensitive function of metallicity, as well as mass \citep{2019ApJ...875...61M}, but there is insufficient data to characterise this function. As a result, initial binary parameter functions derived in the local Universe are typically applied at all metallicities.

\subsubsection*{Mass transfer efficiency}

While Roche lobe overflow can be calculated from a pair of stellar models (using a spherical approximation in the 1D model case), the subsequent evolution of the stars will also depend on how efficiently that mass is transfered to the binary companion as opposed to being ejected from the system. The efficiency is likely stellar mass dependent, and may also be sensitive to other parameters such as metallicity and magnetic fields (which no current population models consider). Since mass transfer is also associated with angular momentum transfer between stars, the prescription used here will also interact with treatment of rotational mixing and also tidal interactions between stars \citep[again often neglected in current models, although see][Chrimes et al., in prep]{2013ApJ...764..166D}.

\subsubsection*{Common Envelope Evolution}

The common envelope evolution phase has a dramatic effect on the evolution of a binary, stripping stellar envelopes and hardening the binaries into very close systems with strong gravitational and tidal interactions. However this phase falls firmly into the category of rapid processes with an extreme dynamic range of scales. As already discussed, these present a challenge to all stellar models. Common envelope phases are usually approximated in evolution models using analytic prescriptions informed by observational constraints and by the few detailed smooth particle hydrodynamics models that attempt to follow this process \citep[see][for a review]{2013A&ARv..21...59I}.

\subsubsection*{Accretion-related emission}

In addition to processing binary models through nebular gas and dust, as in the single star case, a potentially important additional emission component arises directly from the stellar population but is not captured by the stellar spectral synthesis. Stellar populations naturally give rise to binaries in which a compact remnant left after the death of the primary undergoes episodic accretion from the secondary. This can contribute a very high luminosity, very hot spectral component, which is short-lived in each individual binary but can nonetheless contribute significant hard UV and X-ray flux when averaged across a population. Again, this component has not been considered to date by any detailed binary population and spectral synthesis model.

\vspace*{12pt}

\noindent\fbox{\parbox{\textwidth}{
\vspace*{6pt}{\bf Important:}

Before leaving the issue of population synthesis uncertainties, there is an important point to be made: modelling stellar populations with single star evolution tracks {\bf does not remove} the uncertainties associated with binary evolution. Instead it makes an implicit assumption {\bf that we know to be wrong}: that the binary fraction is zero at all masses. Populations constructed from binary tracks {\bf have different luminosities and spectral features} than single star populations at the same mass and age. Thus while any given binary model may be making one or more faulty assumptions, {\bf all} single star models are {\it known} be making faulty assumptions which will lead to {\bf incorrect interpretation} of observational data. The offset between the single star model and the `true' model is likely a function of age, stellar mass, stellar metallicity, gas and dust properties. It is small in the high metallicity, old stellar population regime common in the local Universe, but cannot be neglected at high redshift or in extreme galaxies nearby.
  \vspace*{6pt}}
}

\vspace*{12pt}

\begin{figure}[tb]
\begin{center}
 \includegraphics[width=4.5in]{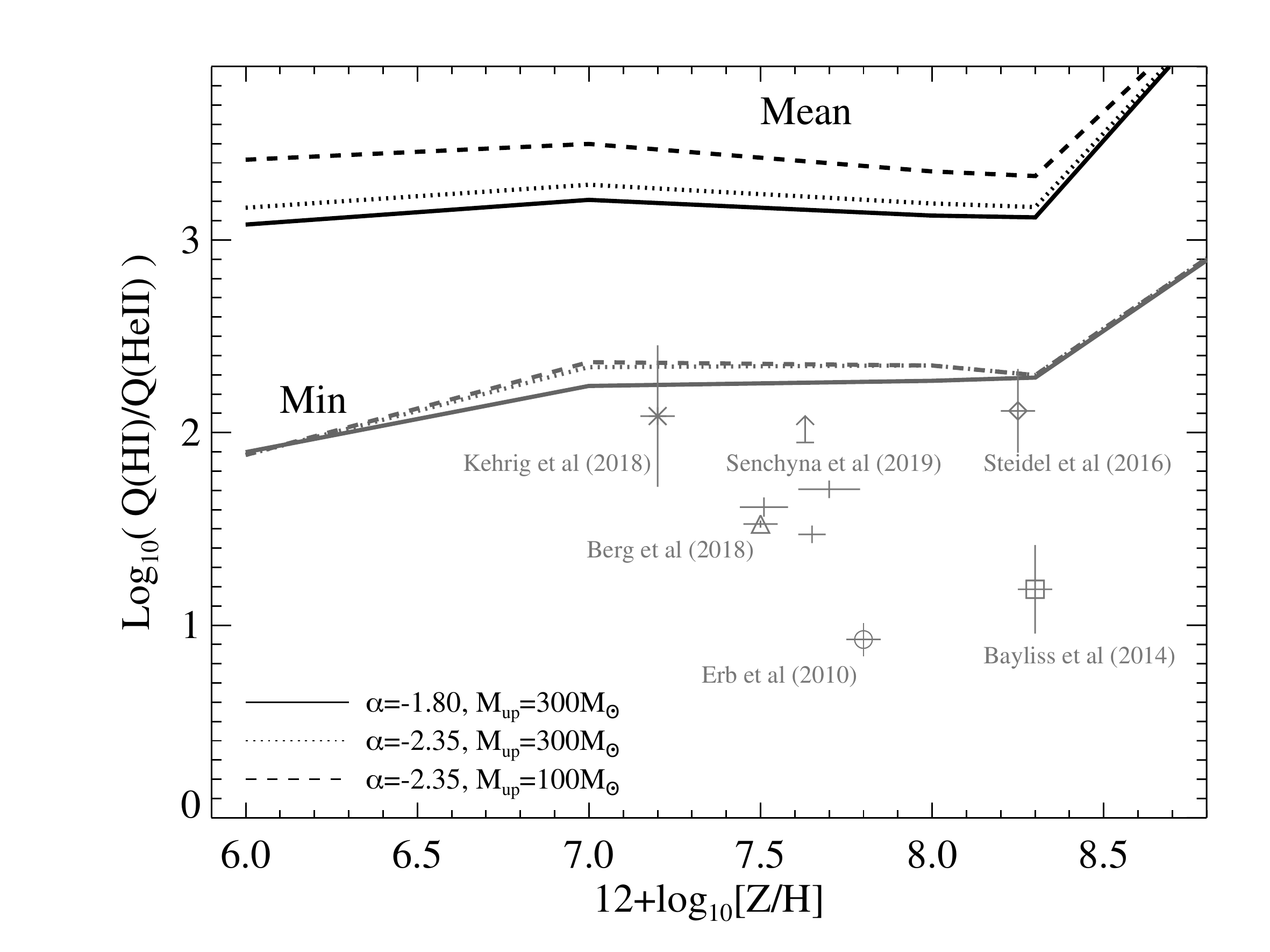} 
 \caption{The hardness of the ionizing spectrum originating from a young starburst galaxy, as measured by the ratio of photons capable of ionizing hydrogen and helium. Lines indicate the minimum and mean photon flux ratios obtained in binary stellar population models at ages up to $10^8$ years, with three different presumed initial mass functions. Points are calculated from observational data for extreme ionizing galaxies in the references labelled. Adapted from Stanway \& Eldridge (2019).}
   \label{fig:qrat}
\end{center}
\end{figure}

\section{Binary Population and Spectral Synthesis (BPASS)}

The Binary Population and Spectral Synthesis (BPASS) project\footnote{See {\tt bpass.auckland.ac.uk} or {\tt warwick.ac.uk/bpass} for full data release and details.} is, as the name suggests, a grid of synthesis models which generates both population and spectral data products for simple stellar populations incorporating binary evolution pathways. The inputs are a custom grid of detailed 1D stellar evolution models generated using the BPASS stellar evolution code \citep[a binary variant of the older STARS code, ][]{1971MNRAS.151..351E}. These are combined with the binary stellar population parameters of \citet{2017ApJS..230...15M}, a supernova kick prescription \citep[we currently use that of][]{2005MNRAS.360..974H}, and a compact-binary gravitational wave evolution algorithm. The models are combined with a high resolution grid of stellar atmospheres for normal stars, together with the Potsdam Wolf Rayet models for stripped stars \citep[PoWR,][]{2003A&A...410..993H,2006A&A...457.1015H,2012A&A...538A..40G} and a custom set of hot O-star models, interpolating the atmosphere models in luminosity, surface gravity and metallicity where required.  Populations are generated at 13 metallicities, with nine initial mass functions in the current distribution.  

The key physical inputs into the v2.1 BPASS stellar evolution and synthesis models are described in full in \citet{2017PASA...34...58E}. Refinements to the binary prescription and treatment of low mass stars resulted in the current v2.2 distribution, which is described in \citet{2018MNRAS.479...75S}. 

Initialy developed as stellar models for supernova progenitors \citep[e.g.][]{2008MNRAS.384.1109E,2013MNRAS.436..774E} and as a population synthesis tool for interpreting distant galaxies \citep{2009MNRAS.400.1019E,2012MNRAS.419..479E}, BPASS has also been applied to nearby massive stellar populations \citep[e.g.][]{2016MNRAS.457.4296W}, old stellar populations \citep[e.g.][]{2018MNRAS.479...75S} and cosmic reionization \citep[e.g.][]{2016MNRAS.456..485S,2015MNRAS.453..960M}. Recent development work has considered the predicted rate and properties of gravitational wave transients \citep[e.g.][]{2019MNRAS.482..870E}, explored gamma-ray burst progenitors (Chrimes et al, in prep), and also returned to the initial focus on low metallicity, young star forming galaxies in both the distant Universe and the rare local extreme galaxy population (see next section).

\section{The Mystery of the Hard Ionizing Spectra}

Amongst the most significant effects of incorporating binary pathways in a stellar population is the production of longer lived, hotter stars than are seen in single star populations. This leads to a harder and stronger rest-frame far ultraviolet radiation field being inferred from the (observable) near ultraviolet. This prediction of far-UV flux is important for calculations of the reionization of the Universe, in which the ionizing photon production efficiency, $\xi_\mathrm{ion}$, is a key parameter \citep[see e.g.][]{2016MNRAS.456..485S}. It is also key to interpreting the nebular emission spectra of distant galaxies - an area which is currently challenging even binary population synthesis models.

\subsection{Observational Evidence}

Analysis of the ionizing spectra of distant galaxies is undertaken primarily through study of those ionizing photons reprocessed by nebular gas in the interstellar medium. Detection of multiple, high signal-to-noise absorption and emission features is now routinely possible in galaxies at $z\sim2-4$ \citep[e.g.][]{2018ApJ...860...75D,2016ApJ...826..159S,2017A&A...608A...4M}, and even higher redshifts in rare cases, a situation which has been markedly improved by the advent of near-infrared sensitive multi-object spectrographs on some of the world's largest telescopes. As a result it is now possible to study spectral line ratios and construct photoionization models of the physical conditions in these star-forming systems. While there is always degeneracy between assumed gas conditions and the spectral energy distribution of the ionizing source, a large number of distant systems are now showing evidence for the existence of a very hard ionizing radiation field. This is observed both in rest-optical line ratios such as [O\,III\,$\lambda$5007\AA]/[O\,II\,$\lambda$3727\AA] or [O\,III\,$\lambda$5007\AA]/H\,$\beta$\,$\lambda$4681\AA,  and directly in the strength of nebular emission lines including C\,III]\,$\lambda$\,1909\AA\ (which requires a 48\,eV ionization potential) and He\,II\,$\lambda$\,1640\AA\ (at 54.4\,eV) in the ultraviolet \citep{2018ApJ...860...75D,2010ApJ...719.1168E,2018ApJ...859..164B}.

  Such work is not limited to the very distant Universe. A large and somewhat disparate class of low redshift ultraviolet luminous and extreme emission line galaxies has now been identified as having similar physical conditions to galaxies in the distant Universe \citep[see e.g.][]{2005ApJ...619L..35H,2016MNRAS.459.2591G,2015A&A...578A.105A}. Of these a subset are `Lyman-continuum leakers', in which a measurable fraction of their flux just shortwards of 912\,\AA\ can be measured directly \cite[e.g.][]{2018MNRAS.478.4851I}. Many of these sources also show evidence for exceptionally hard ionizing spectra, inferred from their nebular emission.

  One measure of this spectral hardness can be provided by the ratio of the recombination lines of hydrogen (typically the H\,I Balmer lines at 4861 or 6563\AA) and helium (He\,II at 1640 or 4686\AA). These are only weakly sensitive to the temperature and electron density of the nebular gas and so can be converted to a ratio of ionizing photon flux: {Q(H\,I)}/{Q(He\,II)} \citep[see][for details]{2019A&A...621A.105S}. Figure \ref{fig:qrat} indicates the inferred photon flux ratios for an indicative sample of extreme emission line sources observed at both high and low redshift: SL2S\,J021737-051329 at $z=1.8$ \citep{2018ApJ...859..164B}, Q2343-BX418 \citep[$z=2.3$,][]{2010ApJ...719.1168E}, SGAS\,J105039.6+001730 \citep[$z=3.6$,][]{2014ApJ...790..144B}, SBS\,0335-052E \citep[$D=54$\,Mpc,][]{2018MNRAS.480.1081K}, HS1442+4250, J0940+2935, J119+5130 and UM133 \citep[$D=11$,\,8,\,22 and 29\,Mpc respectively,][]{2019MNRAS.488.3492S} and the mean properties of stacked $z\sim2.3$ Lyman break galaxies \citep{2016ApJ...826..159S}. While the `typical' distant galaxy is  less extreme than some individual examples, it is clear that very low {Q(H\,I)}/{Q(He\,II)} ratios must be possible in some cases, implying exceptionally hard spectra ionizing the nebular gas.

\subsection{IMF variation}

IMF variations can substantially affect the ionizing flux. An IMF which increases the fraction of massive stars in a population will boost the ionizing flux below both the $\lambda = 912$\,\AA\ hydrogen and  $227.9$\,\AA\ helium ionization edges, with the ratio between them sensive to the slope of the IMF in the massive and very massive star regime. The He\,II line in particular is also sensitive to the upper mass cut-off in the stellar population, particularly at low metallicity and log(age/years)=6.3-6.5.  At these ages very massive stars are approaching the end of their life, swelling to become very luminous giants and (when stripped by winds or binary interactions) Wolf-Rayet stars. A single such star can dominate the emission of a stellar population. The presence or absence of these stars is strongly dependent on the initial mass distribution and how well the massive star IMF is sampled. 

\citet{2019A&A...621A.105S} explored a range of plausible initial mass functions in the BPASS stellar population model formalism, varying both the IMF slope ($\alpha$) and its upper mass cut-off (M$_{up}$). The ionizing photon flux ratio of three representative models are shown overplotted on figure \ref{fig:qrat}. A slight subtlety arises in the direct comparison to the observations, as the latter are calibrated in terms of oxygen abundance while the stellar models are driven by the iron abundance. The models have been offset accordingly to account for a 0.3 dex abundance enhancement in oxygen \citep[e.g.][]{2016ApJ...826..159S}. As the figure demonstrates, while binary models can reproduce the high He\,II line strengths of some galaxies, including that typical of star-forming galaxies at $z\sim2-3$, the exceptionally low photon flux ratios inferred from some extreme galaxies lie an order of magnitude below those inferred from models, regardless of IMF. Indeed, even a synthetic population comprised entirely of very massive 300\,M$_\odot$ stars would not produce the observed line ratios. As discussed in \citet{2019A&A...621A.105S} and \citet{2017PASA...34...58E}, reproducing the hardest ionizing spectra will require exploring ionizing sources besides stellar photospheres.

\begin{figure}[tb]
\begin{center}
 \includegraphics[width=4.5in]{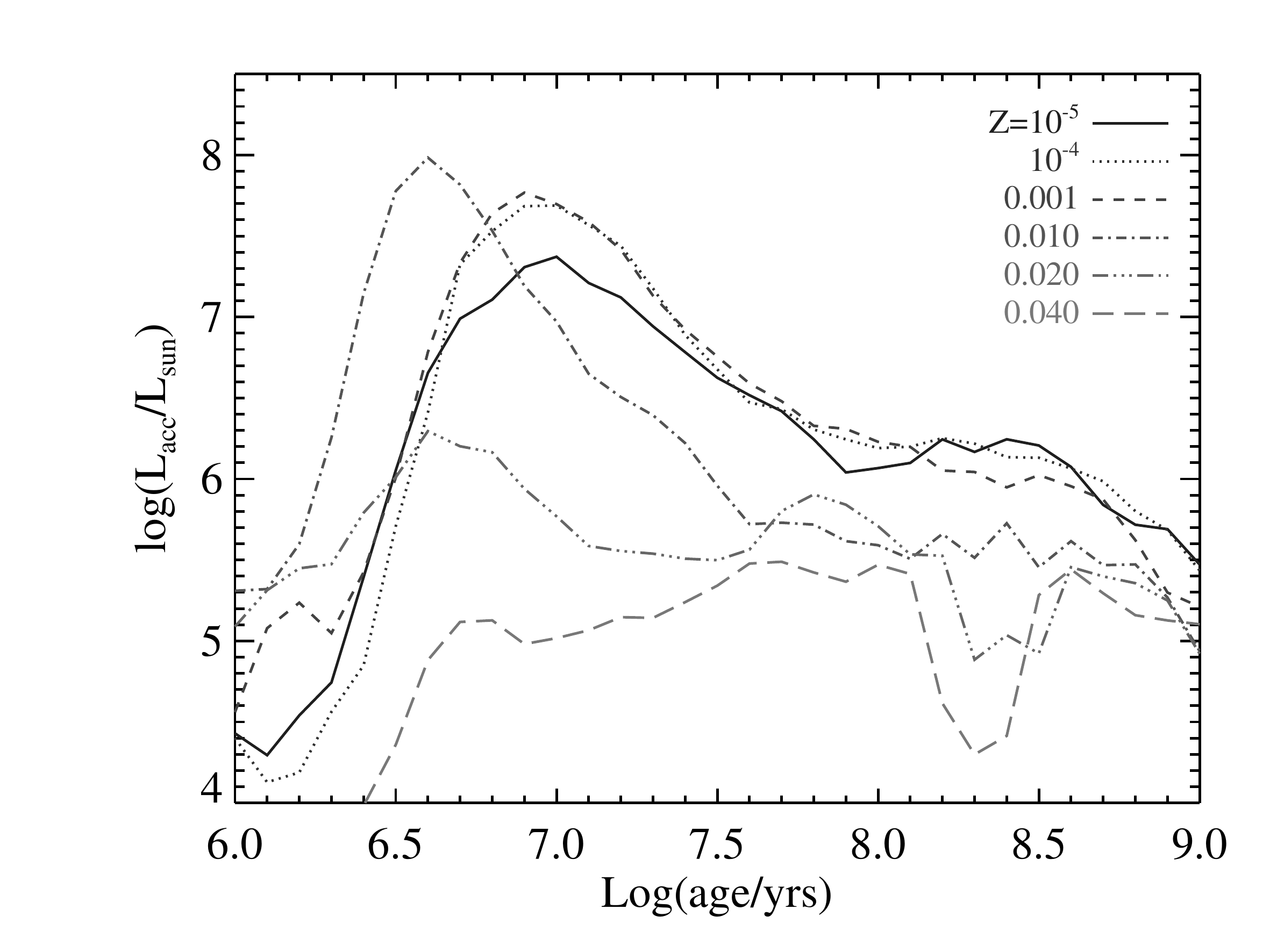} 
 \caption{The evolution of bolometric accretion-powered luminosity from X-ray binaries in BPASS v2.2.1 models, illustrating the different behaviour as a function metallicity. Calibrating these preliminary estimates and the spectral shape of XRB emission is a work in progress (Eldridge et al, in prep; Stanway et al, in prep).}
   \label{fig:xrbs}
\end{center}
\end{figure}

\subsection{X-Ray Binaries}

Accreting binaries are a natural product of binary stellar population synthesis. The potentially important role of accreting compact objects and their associated accretion disks in ionizing photon production has been recognised for a long time \citep[e.g.][]{2000A&A...358..462V}, and has been discussed in a high redshift context at conferences in recent years\footnote{see e.g. {\tt http://www.iastro.pt/research/conferences/lyman2018/pdfs/ElizabethStanway\_Lyman2018.pdf}}, motivated in part by the observational evidence that local Lyman break analogue galaxies tend to show an excess of resolvable X-ray binaries \citep{2013ApJ...774..152B}. Recently \citet{2019A&A...622L..10S} demonstrated that incorporating an ad hoc prescription for the X-ray emission from accreting binaries, informed by observations of extreme local sources, alongside the stellar emission generated from population synthesis models may help to explain the strength of observed He\,II emission in some low metallicity galaxies. 

Limited information on the X-ray binary population generated by BPASS models has been available as an unverified, `on request' data product since the BPASS v2.1 release in 2017 \citep{2017PASA...34...58E}. In Figure \ref{fig:xrbs} we illustrate the time evolution of the accretion-powered luminosity component calculated by BPASS v2.2.1 \citep{2018MNRAS.479...75S} as a function of metallicity. Tracks assume a 10\% efficiency in conversion of gravitational potential energy to luminosity, and give the average population bolometric luminosity in each time bin for a simple stellar population with initial mass of $10^6$\,M$_\odot$. Models are smoothed across three time bins to account for stochasticity introduced by the underlying grid of stellar models. The figure illustrates the key motivation for exploring X-ray binaries as a source of ionizing photons at high redshift: low metallicity stars have weaker winds, they retain more of their initial mass at the end of their lifetime, form more massive compact remnants, and so are more likely to form highly luminous high mass X-Ray Binaries. Rather than the hundreds of millions of years it takes for higher metallicity, lower mass X-ray binaries to form, low metallicity systems establish a substantial accretion luminosity within a few Myr of the onset of star formation, and remains luminous for almost a Gyr.

Analysis to model the accretion disk dominated emission spectrum and calibrate the efficiency of these systems against observational evidence is ongoing. We expect the next BPASS data release to include calibrated accretion-powered ionizing spectra and data on the X-ray binary population, derived consistently with the underlying stellar population and integrated stellar spectra (Eldridge et al, in prep; Stanway et al in prep.)

\section{Final Thoughts}

The evolution of population and spectral synthesis models has been rapid and dramatic in recent years. While in part driven by advances in computational capacity, perhaps more important has been the development and implementation of prescriptions for ever more complex aspects of stellar evolution, and the observational instrument developments which have informed these. The recognition that distant galaxies provide a test bed for modelling the behaviours of high mass, low metallicity stellar populations has formed a virtuous circle with the growing realisation that such populations cannot be neglected when interpreting the distant Universe. Local exemplars of extreme galaxies providing opportunities for more detailed studies, an opportunity which is being widely exploited by the extragalactic community, and which has begun to be realised by the stellar evolution community in turn.

The contribution of ALMA has not been discussed in detail in this work. This should not be taken to indicate that it does not play a role. Long baseline imaging, particularly of lensed arcs, has begun to explore the physical conditions in distant star forming galaxies in exquisite detail and to redshifts inconceivable just a few years ago \citep[e.g.][]{2018Natur.553..178S,2018Natur.557..392H,2019PASJ..tmp...70H}. Intriguingly, one result of such observations has been to identify higher than expected ratios between the [O\,III] 88$\mu$m and [C\,II] 158$\mu$m emission lines in such sources, and potentially also elevated dust temperatures. While interpreting these features is challenging given the observational constraints involved, both hint at the same hard ionizing spectrum and excess of hot photons implied by the He\,II observations in the rest-frame ultraviolet and optical. Just as importantly, ALMA observations may be key to resolving some of the uncertainties in massive star evolution, particularly in terms of the mass loss rates and outflows from giant stars observed locally.

The launch of the James Webb Space Telescope presents both an opportunity and a challenge for the population synthesis and binary modelling communities. The large primary mirror of JWST will enable resolved stellar population studies to be extended to far greater distances than currently possible with Hubble or ground-based imaging. This will bring more low metallicity and young, massive star forming regions within reach, and enable better characterisation of close binary populations and their interactions.  On the other hand, JWST's infrared sensitivity range is not optimal for exploring the hottest stars or their ionization regions nearby, for which a future ultraviolet mission will be needed. Where JWST really does promise to revolutionize our understanding is in the rest-ultraviolet and rest-optical spectral characteristics of distant galaxies. Spectroscopic observations currently carried out with great difficulty and expense in telescope time from the ground, have the potential to become routine with the massively multiplexed, near-to-mid infrared spectrographs mounted on JWST (NIRSPEC and MIRI), and to be pushed to far higher redshift. At the same time, improved infrared (rest-optical) photometry will allow the low mass stellar populations of these galaxies to be better characterised by spectral energy distribution fitting than is currently possible.

\begin{figure}[tb]
\begin{center}
 \includegraphics[width=5.5in]{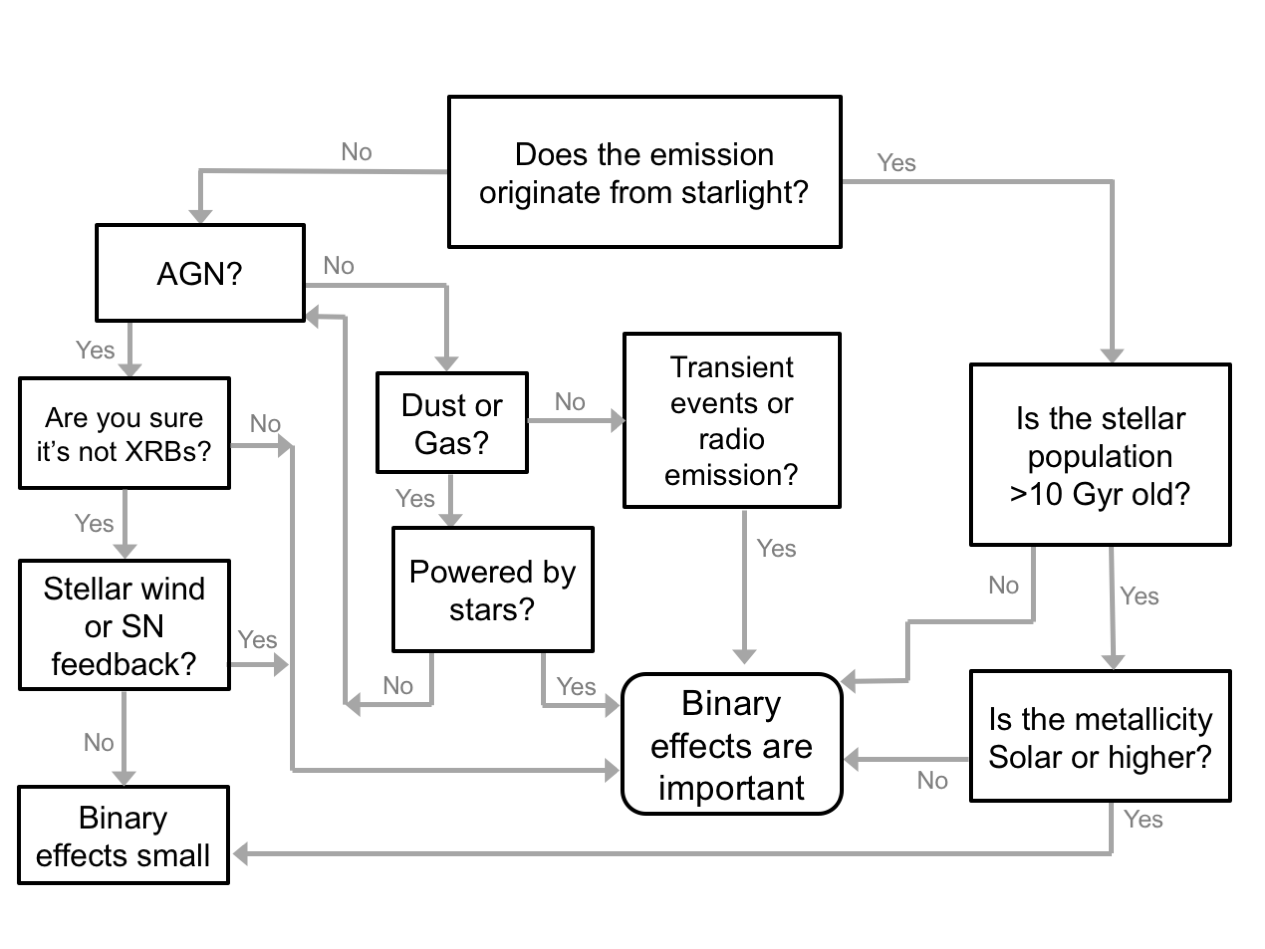} 
 \caption{A non-exhaustive decision tree indicating regions of observational parameter space in which considering the effects of stellar binaries may have a significant effect on the interpretation of results.}
   \label{fig:binaries}
\end{center}
\end{figure}

On the other hand, the expected improvement in signal-to-noise of observations of these extreme systems is likely to bring into sharp relief the limitations and uncertainties in existing stellar and spectral synthesis models. It is likely that the modelling effort will evolve hand-in-hand with the insights yielded by observational data, as some model parameters are ruled out, and others adjusted. Crucially, this effort will require an understanding of the role and impact of binary star evolution. As Figure \ref{fig:binaries} illustrates, we are now firmly within regimes in which binary evolution pathways can significantly affect the interpretation of stellar populations. Neglecting them would be a collective act of  voluntary myopia on the part of the community akin to wearing dark and clouded glasses to an exhibition of fine art - while the broadest strokes might be visible, we would be ignorant of the exquisite detail, intellectual, and indeed emotional, stimulation which the complex Universe we inhabit has to offer.

\section*{Acknowledgements}

ERS acknowledges support from the University of Warwick, and under UK STFC consolidated grant ST/P000495/1. JJE acknowledges support fom the University of Auckland and also the Royal Society Te Ap\={a}rangi of New Zealand under Marsden Fund grant UOA1818.


\begin{thebibliography}{}

\bibitem[Amor{\'\i}n et al.(2015)]{2015A&A...578A.105A} Amor{\'\i}n, R., P{\'e}rez-Montero, E., Contini, T., et al.\ 2015, \aap, 578, A105
\bibitem[Basu-Zych et al.(2013)]{2013ApJ...774..152B} Basu-Zych, A.~R., Lehmer, B.~D., Hornschemeier, A.~E., et al.\ 2013, \apj, 774, 152
\bibitem[Bayliss et al.(2014)]{2014ApJ...790..144B} Bayliss, M.~B., Rigby, J.~R., Sharon, K., et al.\ 2014, \apj, 790, 144
\bibitem[Belczynski et al.(2008)]{2008ApJS..174..223B} Belczynski, K., Kalogera, V., Rasio, F.~A., et al.\ 2008, \apjs, 174, 223
\bibitem[Berg et al.(2018)]{2018ApJ...859..164B} Berg, D.~A., Erb, D.~K., Auger, M.~W., et al.\ 2018, \apj, 859, 164
\bibitem[Bruzual, \& Charlot(2003)]{2003MNRAS.344.1000B} Bruzual, G., \& Charlot, S.\ 2003, \mnras, 344, 1000
\bibitem[Conroy(2013)]{2013ARA&A..51..393C} Conroy, C.\ 2013, \araa, 51, 393
\bibitem[da Cunha et al.(2008)]{2008MNRAS.388.1595D} da Cunha, E., Charlot, S., \& Elbaz, D.\ 2008, \mnras, 388, 1595
\bibitem[De Donder, \& Vanbeveren(2004)]{2004NewAR..48..861D} De Donder, E., \& Vanbeveren, D.\ 2004, New Astronomy Review, 48, 861
\bibitem[de Mink et al.(2013)]{2013ApJ...764..166D} de Mink, S.~E., Langer, N., Izzard, R.~G., et al.\ 2013, \apj, 764, 166
\bibitem[Du et al.(2018)]{2018ApJ...860...75D} Du, X., Shapley, A.~E., Reddy, N.~A., et al.\ 2018, \apj, 860, 75
\bibitem[Dufton et al.(2019)]{2019A&A...626A..50D} Dufton, P.~L., Evans, C.~J., Hunter, I., et al.\ 2019, \aap, 626, A50
\bibitem[Eggleton(1971)]{1971MNRAS.151..351E} Eggleton, P.~P.\ 1971, \mnras, 151, 351
\bibitem[Eldridge et al.(2008)]{2008MNRAS.384.1109E} Eldridge, J.~J., Izzard, R.~G., \& Tout, C.~A.\ 2008, \mnras, 384, 1109
\bibitem[Eldridge \& Stanway(2009)]{2009MNRAS.400.1019E} Eldridge, J.~J., \& Stanway, E.~R.\ 2009, \mnras, 400, 1019
\bibitem[Eldridge \& Stanway(2012)]{2012MNRAS.419..479E} Eldridge, J.~J., \& Stanway, E.~R.\ 2012, \mnras, 419, 479
\bibitem[Eldridge et al.(2013)]{2013MNRAS.436..774E} Eldridge, J.~J., Fraser, M., Smartt, S.~J., et al.\ 2013, \mnras, 436, 774
\bibitem[Eldridge et al.(2017)]{2017PASA...34...58E} Eldridge, J.~J., Stanway, E.~R., Xiao, L., et al.\ 2017, \pasa, 34, e058
\bibitem[Eldridge et al.(2019)]{2019MNRAS.482..870E} Eldridge, J.~J., Stanway, E.~R., \& Tang, P.~N.\ 2019, \mnras, 482, 870
\bibitem[Erb et al.(2010)]{2010ApJ...719.1168E} Erb, D.~K., Pettini, M., Shapley, A.~E., et al.\ 2010, \apj, 719, 1168
\bibitem[Ferland et al.(1998)]{1998PASP..110..761F} Ferland, G.~J., Korista, K.~T., Verner, D.~A., et al.\ 1998, \pasp, 110, 761
\bibitem[G{\"o}tberg et al.(2018)]{2018A&A...615A..78G} G{\"o}tberg, Y., de Mink, S.~E., Groh, J.~H., et al.\ 2018, \aap, 615, A78
\bibitem[Gr{\"a}fener et al.(2012)]{2012A&A...538A..40G} Gr{\"a}fener, G., Owocki, S.~P., \& Vink, J.~S.\ 2012, \aap, 538, A40
\bibitem[Greis et al.(2016)]{2016MNRAS.459.2591G} Greis, S.~M.~L., Stanway, E.~R., Davies, L.~J.~M., et al.\ 2016, \mnras, 459, 2591
\bibitem[Hamann \& Gr{\"a}fener(2003)]{2003A&A...410..993H} Hamann, W.-R., \& Gr{\"a}fener, G.\ 2003, \aap, 410, 993
\bibitem[Hamann et al.(2006)]{2006A&A...457.1015H} Hamann, W.-R., Gr{\"a}fener, G., \& Liermann, A.\ 2006, \aap, 457, 1015
\bibitem[Han et al.(2007)]{2007MNRAS.380.1098H} Han, Z., Podsiadlowski, P., \& Lynas-Gray, A.~E.\ 2007, \mnras, 380, 1098
\bibitem[Hashimoto et al.(2018)]{2018Natur.557..392H} Hashimoto, T., Laporte, N., Mawatari, K., et al.\ 2018, \nat, 557, 392
\bibitem[Hashimoto et al.(2019)]{2019PASJ..tmp...70H} Hashimoto, T., Inoue, A.~K., Mawatari, K., et al.\ 2019, \pasj, 70
\bibitem[Heckman et al.(2005)]{2005ApJ...619L..35H} Heckman, T.~M., Hoopes, C.~G., Seibert, M., et al.\ 2005, \apjl, 619, L35
\bibitem[Hobbs et al.(2005)]{2005MNRAS.360..974H} Hobbs, G., Lorimer, D.~R., Lyne, A.~G., et al.\ 2005, \mnras, 360, 974
\bibitem[Hopkins(2018)]{2018PASA...35...39H} Hopkins, A.~M.\ 2018, \pasa, 35, 39
\bibitem[Hurley et al.(2002)]{2002MNRAS.329..897H} Hurley, J.~R., Tout, C.~A., \& Pols, O.~R.\ 2002, \mnras, 329, 897
\bibitem[Ivanova et al.(2013)]{2013A&ARv..21...59I} Ivanova, N., Justham, S., Chen, X., et al.\ 2013, \aapr, 21, 59
\bibitem[Izotov et~al. (2018)]{2018MNRAS.478.4851I}{Izotov}, Y.~I., {Worseck}, G., {Schaerer}, D., {et~al.} 2018, \mnras, 478, 4851
\bibitem[Kehrig et al.(2018)]{2018MNRAS.480.1081K} Kehrig, C., V{\'\i}lchez, J.~M., Guerrero, M.~A., et al.\ 2018, \mnras, 480, 1081
\bibitem[Leitherer et al.(1999)]{1999ApJS..123....3L} Leitherer, C., Schaerer, D., Goldader, J.~D., et al.\ 1999, \apjs, 123, 3
\bibitem[Leitherer et al.(2014)]{2014ApJS..212...14L} Leitherer, C., Ekstr{\"o}m, S., Meynet, G., et al.\ 2014, \apjs, 212, 14
\bibitem[Ma et al.(2015)]{2015MNRAS.453..960M} Ma, X., Kasen, D., Hopkins, P.~F., et al.\ 2015, \mnras, 453, 960
\bibitem[Madau et al.(1996)]{1996MNRAS.283.1388M} Madau, P., Ferguson, H.~C., Dickinson, M.~E., et al.\ 1996, \mnras, 283, 1388
\bibitem[Maraston(2005)]{2005MNRAS.362..799M} Maraston, C.\ 2005, \mnras, 362, 799
\bibitem[Maseda et al.(2017)]{2017A&A...608A...4M} Maseda, M.~V., Brinchmann, J., Franx, M., et al.\ 2017, \aap, 608, A4
\bibitem[Moe \& Di Stefano(2017)]{2017ApJS..230...15M} Moe, M., \& Di Stefano, R.\ 2017, \apjs, 230, 15
\bibitem[Moe et al.(2019)]{2019ApJ...875...61M} Moe, M., Kratter, K.~M., \& Badenes, C.\ 2019, \apj, 875, 61
\bibitem[Nhung et al.(2019)]{2019arXiv190803311N} Nhung, P.~T., Hoai, D.~T., Tuan-Anh, P., et al.\ 2019, arXiv e-prints, arXiv:1908.03311
\bibitem[Pols et al.(1995)]{1995MNRAS.274..964P} Pols, O.~R., Tout, C.~A., Eggleton, P.~P., et al.\ 1995, \mnras, 274, 964
\bibitem[Rosenfield et al.(2014)]{2014ApJ...790...22R} Rosenfield, P., Marigo, P., Girardi, L., et al.\ 2014, \apj, 790, 22
\bibitem[Sana et al.(2012)]{2012Sci...337..444S} Sana, H., de Mink, S.~E., de Koter, A., et al.\ 2012, Science, 337, 444
\bibitem[Schaerer et al.(2019)]{2019A&A...622L..10S} Schaerer, D., Fragos, T., \& Izotov, Y.~I.\ 2019, \aap, 622, L10
\bibitem[Schneider et al.(2018a)]{2018A&A...618A..73S} Schneider, F.~R.~N., Ram{\'\i}rez-Agudelo, O.~H., Tramper, F., et al.\ 2018, \aap, 618, A73
\bibitem[Schneider et al.(2018b)]{2018Sci...359...69S} Schneider, F.~R.~N., Sana, H., Evans, C.~J., et al.\ 2018, Science, 359, 69
\bibitem[Senchyna et al.(2019)]{2019MNRAS.488.3492S} Senchyna, P., Stark, D.~P., Chevallard, J., et al.\ 2019, \mnras, 488, 3492
\bibitem[Shapley et al.(2003)]{2003ApJ...588...65S} Shapley, A.~E., Steidel, C.~C., Pettini, M., et al.\ 2003, \apj, 588, 65
\bibitem[Smit et al.(2018)]{2018Natur.553..178S} Smit, R., Bouwens, R.~J., Carniani, S., et al.\ 2018, \nat, 553, 178
\bibitem[Stanway et al.(2016)]{2016MNRAS.456..485S} Stanway, E.~R., Eldridge, J.~J., \& Becker, G.~D.\ 2016, \mnras, 456, 485
\bibitem[Stanway \& Eldridge(2018)]{2018MNRAS.479...75S} Stanway, E.~R., \& Eldridge, J.~J.\ 2018, \mnras, 479, 75
\bibitem[Stanway \& Eldridge(2019)]{2019A&A...621A.105S} Stanway, E.~R., \& Eldridge, J.~J.\ 2019, \aap, 621, A105
\bibitem[Steidel et al.(2016)]{2016ApJ...826..159S} Steidel, C.~C., Strom, A.~L., Pettini, M., et al.\ 2016, \apj, 826, 159
\bibitem[Toonen, \& Nelemans(2013)]{2013A&A...557A..87T} Toonen, S., \& Nelemans, G.\ 2013, \aap, 557, A87
\bibitem[Topping \& Shull(2015)]{2015ApJ...800...97T} Topping, M.~W., \& Shull, J.~M.\ 2015, \apj, 800, 97
\bibitem[van Bever \& Vanbeveren(1998)]{1998A&A...334...21V} van Bever, J., \& Vanbeveren, D.\ 1998, \aap, 334, 21
\bibitem[Van Bever \& Vanbeveren(2000)]{2000A&A...358..462V} Van Bever, J., \& Vanbeveren, D.\ 2000, \aap, 358, 462
\bibitem[Willems \& Kolb(2004)]{2004A&A...419.1057W} Willems, B., \& Kolb, U.\ 2004, \aap, 419, 1057
\bibitem[Wofford et al.(2016)]{2016MNRAS.457.4296W} Wofford, A., Charlot, S., Bruzual, G., et al.\ 2016, \mnras, 457, 4296
\bibitem[Zhang et al.(2005)]{2005MNRAS.364..503Z} Zhang, F., Li, L., \& Han, Z.\ 2005, \mnras, 364, 503

\vspace{24pt}
  



\end{thebibliography}
\end{document}